\documentclass[12pt, A4]{article}
\usepackage{amssymb}
\usepackage{amsmath}
\usepackage{amsfonts}
\usepackage[dvips]{graphicx}
\usepackage{setspace}
\usepackage{latexsym}
\usepackage{epsf}
\usepackage{epsfig}
\usepackage{color}
\usepackage{hyperref}
\hypersetup{
    colorlinks=true, 
    linktoc=all,     
    linkcolor=black,  
    pdfstartview={XYZ null null 1.00},
}

\begin{document}
\vspace*{\fill}
\begin{center}
\textbf{\Large A Cosmological Solution to Mimetic Dark Matter}
\end{center} 
\vspace{0.2 cm}
\begin{center}
\textbf{Hassan Saadi} \\
\vspace{0.1 cm}
\textit{Physics Department, American University of Beirut, Lebanon}
\end{center}
\vspace{1.5 cm}
\begin{abstract}
\vspace{0.3 cm}
In this paper, a cosmological solution to Mimetic Dark Matter (MDM) for an exponential potential is provided. Then, a solution for the $0-i$ perturbed Einstein's differential equation of MDM is obtained based on an exponential potential that satisfies inflation for some initial conditions. Another general potential is suggested that incorporates inflation too. Then, quantum perturbations are included. The constants in the model can be tuned to be in agreement with the amplitude fluctuation of the cosmic microwave background (CMB) radiation. Finally, the spectral index is calculated for the suggested potentials. Moreover, MDM is shown to be a viable model to produce dark matter, inflation, and CMB's fluctuation. 
\end{abstract}
\vspace*{\fill}
\clearpage
\section{Introduction}
A modification of general relativity was proposed in \cite{MDM} where a metric $g_{\mu\nu}$ is defined by a scalar field $\phi$ and an auxiliary metric $\tilde{g}_{\mu\nu}$
\begin{equation}
g_{\mu\nu}=(\tilde{g}^{\alpha\beta} \partial_{\alpha}\phi \partial_{\beta}\phi) \tilde{g}_{\mu\nu}
\end{equation}
The equations of motion that result are similar to Einstein's equations of motion with an extra mode term that mimics cold dark matter even in the absence of normal matter. Mimetic dark matter is an interesting model because it's a model that works not only on a cosmological scale, but also a model that works on a galactic scale after adding higher derivative terms that alter the speed of sound \cite{smallscale} \cite{cosmology}. For further discussion about mimetic dark matter, degrees of freedom, and extensions check \cite{extension}.\\
\indent Consider the actions in \cite{cosmology} \cite{Lag} \cite{Lag1} \cite{Lag2},
\begin{equation}
S=\int d^{4}x \sqrt{-g}\Big[-\frac{1}{2}R(g_{\mu\nu}) +\lambda(g^{\mu\nu}\partial_\mu \phi \partial_\nu\phi -1)-V(\phi)+ \mathcal{L}_{m}(g_{\mu\nu,...})\Big]
\label{act1}
\end{equation}
where $-\frac{1}{2}R(g_{\mu\nu})$ is the Lagrangian of General Relativity, $V(\phi)$ is a potential, and $\mathcal{L}_{m}$ is the Lagrangian of matter. By varying the action with respect to $g^{\mu\nu}$, $\phi$, and $\lambda$, and taking the trace, equations (13) and (14) in \cite{cosmology} are obtained,
\begin{eqnarray}
G_{\mu\nu} &=& (G-T-4V)\partial_\mu \phi \partial_\nu\phi +g_{\mu\nu}V(\phi)+T_{\mu\nu} \\
0 &=& \frac{1}{\sqrt{-g}}\partial_{\kappa}\Big(\sqrt{-g}g^{\kappa\lambda}\partial_{\lambda}\phi(G-T) \Big) \\
1 &=& g^{\mu\nu}\partial_\mu \phi \partial_\nu\phi \label{normalization}
\end{eqnarray}
where $T_{\mu\nu}$ is the energy-momentum tensor. Note that the normalization condition on the four-velocity $u^{\mu}u_{\mu}=1$ is the normalization condition (\ref{normalization}). \\
\indent For a spatially flat FRW universe with metric,
\begin{equation}
ds^{2}=dt^{2}-a^{2}(t)\delta_{ik}dx^{i}dx^{k}
\end{equation}
taking $\phi=t$ and calculating the time-time component of equation (3), which is the Friedmann equation in the absence of ordinary matter ($T_{\mu\nu}=0$) \cite{cosmology}
\begin{equation}
H^{2}=\frac{1}{a^{3}}\int a^{2} V da
\label{Frie}
\end{equation}
where, 
\begin{equation}
H \equiv \frac{\dot{a}}{a}
\end{equation}
and $a(t)$ is determined by given potential. Note that Mimetic
Dark Matter appears as an integration constant in the right hand
side of (\ref{Frie}) which gives a non-trivial solution even for
$V=0$. By multiplying equation (6) by $a^{3}$ and differentiating with respect to time, and substituting $y=a^{\frac{3}{2}}$,
\begin{equation}
\ddot{y}-\frac{3}{4}V(t)y=0
\label{diffpot}
\end{equation}
This equation allows to find cosmological solutions for $a(t)$ for any given potential.
\section{Solution for Exponential Potential}
Plugging in the exponential potential,
\begin{equation}
V=\alpha e^{-\kappa t}
\label{exppot}
\end{equation} 
where $\alpha$ and $\kappa$ are constants, we obtain
\begin{equation}
\ddot{y}-\frac{3}{4}\alpha e^{-\kappa t}y=0
\label{diffpotexp}
\end{equation}
By applying the transformation $s=\frac{\sqrt{-3\alpha}}{\kappa} e^{\frac{-\kappa t}{2}}$, differential equation (\ref{diffpotexp}) transforms to,
\begin{equation}
s^{2}\frac{d^{2}y}{ds^{2}}+s\frac{dy}{ds}+s^{2}y=0
\end{equation}
The solution to this differential equation is well-known by Bessel's functions,
\begin{equation}
y(t)=C_{1} J_{0}(\frac{\sqrt{-3\alpha}}{\kappa} e^{\frac{-\kappa t}{2}}) + C_{2} Y_{0}(\frac{\sqrt{-3\alpha}}{\kappa} e^{\frac{-\kappa t}{2}})
\end{equation}
where $C_{1}$ and $C_{2}$ are constants. The form of $a(t)$ is,
\begin{equation}
a(t)=[C_{1} J_{0}(\frac{\sqrt{-3\alpha}}{\kappa} e^{\frac{-\kappa t}{2}}) + C_{2} Y_{0}(\frac{\sqrt{-3\alpha}}{\kappa} e^{\frac{-\kappa t}{2}})]^{\frac{2}{3}}
\label{scalefactor}
\end{equation}
It can be deduced that for $t\longrightarrow \infty$, $a(t) \propto t^{\frac{2}{3}}$ which is similar to the scaling factor of a matter-dominated universe. On the other hand, for $t\longrightarrow 0$,
\begin{eqnarray}
y(t)&=& C_{1}e^{\sqrt{\frac{3\alpha}{4}}t}+ C_{2}e^{-\sqrt{\frac{3\alpha}{4}}t} \\
a(t)&=& \Big[C_{1}e^{\sqrt{\frac{3\alpha}{4}}t}+ C_{2}e^{-\sqrt{\frac{3\alpha}{4}}t}\Big]^{\frac{2}{3}}
\end{eqnarray}
For $\alpha$ positive, $a(t)$ grows exponentially as in an inflationary universe. However, for $\alpha$ negative, $a(t)$ leads to an oscillatory universe in the beginning of time. The energy density of mimetic matter can be obtained as
\begin{equation}
\tilde{\rho}=3(\frac{\dot{a}}{a})^{2}=- \alpha e^{-\kappa t}\Big[\frac{C_{1} J_{-1}(\frac{\sqrt{-3\alpha}}{\kappa} e^{\frac{-\kappa t}{2}}) + C_{2} Y_{-1}(\frac{\sqrt{-3\alpha}}{\kappa} e^{\frac{-\kappa t}{2}})}{C_{1} J_{0}(\frac{\sqrt{-3\alpha}}{\kappa} e^{\frac{-\kappa t}{2}}) + C_{2} Y_{0}(\frac{\sqrt{-3\alpha}}{\kappa} e^{\frac{-\kappa t}{2}})}\Big]^{2}
\end{equation}
and the pressure,
\begin{equation}
\tilde{p}= -V(t)= - \alpha e^{-\kappa t}
\end{equation}
and the equation of state is
\begin{equation}
w=\frac{\tilde{p}}{\tilde{\rho}}=  \Big[\frac{C_{1} J_{0}(\frac{\sqrt{-3\alpha}}{\kappa} e^{\frac{-\kappa t}{2}}) + C_{2} Y_{0}(\frac{\sqrt{-3\alpha}}{\kappa} e^{\frac{-\kappa t}{2}})}{C_{1} J_{-1}(\frac{\sqrt{-3\alpha}}{\kappa} e^{\frac{-\kappa t}{2}}) + C_{2} Y_{-1}(\frac{\sqrt{-3\alpha}}{\kappa} e^{\frac{-\kappa t}{2}})}\Big]^{2}
\end{equation}
Moreover, it can be deduced from (\ref{diffpotexp}) that the density, pressure, and equation of state  evolve like dust in a matter-dominated universe for $t\longrightarrow \infty$
\begin{eqnarray}
\tilde{\rho} &=& \frac{4}{3t^{2}} \\
\tilde{p} &=& -V(t)= 0 \\
w &=& 0
\end{eqnarray}
For $t\longrightarrow 0$,
\begin{eqnarray}
\tilde{\rho} &\approx& \alpha \label{e0} \quad \Big(\text{for} \; (C_{1} \gg C_{2})\; \text{and} \; (C_{1} \ll C_{2})\Big)\\
\tilde{\rho} &\approx& 0 \label{e02} \quad (C_{1} \approx C_{2})\\
\tilde{p} &=& -\alpha \label{p0} \\
w &\approx& -1 \label{w0} \quad \Big(\text{for} \; (C_{1} \gg C_{2})\; \text{and} \; (C_{1} \ll C_{2})\Big)\\
w &\approx& 0 \quad \label{w02} (C_{1} \approx C_{2})
\end{eqnarray}
The equation of state for $t\longrightarrow 0$ (\ref{w0}) is at the Phantom Divide Line similar to the equation of state of a positive cosmological constant that drives inflation but without a graceful exit. In order to trigger inflation in the beginning of time, $\ddot{a}(t)>0$ must be satisfied. The acceleration equation is,
\begin{equation}
\frac{\ddot{a}}{a}= - \dfrac{4\pi G}{3}\big(\rho+ 3p\big)
\end{equation}
Hence, $\rho + 3p < 0$ must be true. Density is always positive; therefore, we must have negative pressure satisfying
\begin{equation}
p < - \frac{\rho}{3}
\end{equation}
This is valid for $t$ very small, positive $\alpha$, and all initial conditions $C_{1}$ and $C_{2}$. A 60 e-folds inflation can be generated in this picture for any $\alpha$ because it satisfies the inequality. Let's consider another potential \cite{cosmology}
\begin{equation}
V(t)=\frac{\alpha t^{2n}}{e^{t\kappa}+1} \qquad \text{for $n>-1$}
\label{newpot}
\end{equation}
given that $e^{t\kappa} \gg t^{2n}$ is true always for positive time and suitable $n$. As $t \longrightarrow \infty$ and $t \longrightarrow 0$ it evolves as  $a(t)\propto t^{\frac{2}{3}}$, and as $t \longrightarrow - \infty$ it generates inflation satisfying the 60 e-folds condition with
\begin{equation}
a(t) \propto e^{-\sqrt{\frac{\alpha}{3(n+1)^{2}}}\:t^{2}}
\label{newscalefactor}
\end{equation}
with
\begin{equation}
H=\frac{\dot{a}}{a}=-\sqrt{\frac{\alpha}{3}} \; t^{n} 
\label{newH}
\end{equation}
The number of e-folds is calculated by
\begin{equation}
N= \int_{t_{i}}^{t_{f}} H dt
\label{efolds}
\end{equation}
Note that at $t \longrightarrow \infty$ both potentials (\ref{exppot}) and (\ref{newpot}) behaves the same because at $t \longrightarrow \infty$ (\ref{newpot}) can be approximated as (\ref{exppot}).
In order to give an estimate for $\alpha$, calculate (\ref{efolds}) for 60 e-folds for (\ref{newscalefactor}), and noting that $t_{i}^{2} \gg t_{f}^{2}$ for this model because inflation starts from $-\infty$; and hence 
\begin{equation}
\alpha \simeq \Big(\frac{540(n+1)}{t_{i}^{n+1}}\Big)^{2}
\label{alpha}
\end{equation}
\section{Perturbative Solution of the Scalar Field in the Newtonian Gauge}
Scalar perturbations are considered in the Newtonian gauge. Vector perturbations are neglected because they decay in an expanding universe and because inflation rules out large primordial vector perturbations. In the Newtonian Gauge, the metric of the perturbed universe can be expressed as \cite{Mukhanov}
\begin{equation}
ds^{2}=(1+2\Phi)dt^{2}-(1-2\Phi)a^{2}(t)\delta_{ij}dx^{i}dx^{j}
\end{equation}
and 
\begin{equation}
\phi=t+\delta \phi
\label{phi}
\end{equation} 
is the  perturbation of the scalar field. Perturbing the equations that result from action (\ref{act1}), it can be deduced from \cite{cosmology} that there's one expression for $\delta\phi$ for all wavelengths, 
\begin{eqnarray}
\delta\phi &=& A \frac{1}{a} \int a dt \label{old pert1} \\
\Phi &=& \delta\dot{\phi} = A \Big(1-\frac{H}{a}\int a dt)
\label{old pert2}
\end{eqnarray}
Note that the first equality in (\ref{old pert2}) is deduced from (\ref{normalization}). When spatial derivatives are neglected, expressions (\ref{old pert1}) and (\ref{old pert2}) are exact general solutions for long wavelength cosmological perturbations \cite{Mukhanov}. If $\Phi$ is calculated by using action (\ref{act1}), we would get (\ref{old pert2}) for all wavelengths, and it doesn't distinguish between short and long wavelength perturbations \cite{cosmology}. We wouldn't be able to define quantum perturbations that are short wavelength perturbations. Therefore, in order to account for different wavelengths' perturbations, a term is added to the action $(\ref{act1})$, $\frac{1}{2}\gamma (\Box\phi)^{2}$ where $\gamma$ is a constant and $\Box= g^{\mu \nu}\nabla_{\mu}\nabla_{\nu} $. The action becomes \cite{cosmology},
\begin{equation}
S=\int d^{4}x \sqrt{-g}\Big[-\frac{1}{2}R(g_{\mu\nu}) +\lambda(g^{\mu\nu}\partial_\mu \phi \partial_\nu\phi -1)-V(\phi)+ \frac{1}{2}\gamma (\Box\phi)^{2}\Big]
\label{act2}
\end{equation}  
The $0-0$ and $i-j$ Einstein's equations remain the same up to a normalization constant. On the other hand, the perturbed $0-i$ Einstein's equation \cite{cosmology}
\begin{equation}
\delta\ddot{\phi}+H\delta\dot{\phi}-\frac{c^{2}_{s}}{a^{2}}\Delta\delta\phi+\dot{H}\delta\phi=0
\label{0-i pert}
\end{equation} 
where
\begin{eqnarray}
c^{2}_{s} &=& \frac{\gamma}{2-3\gamma}
\label{sound}
\end{eqnarray}
Note that after adding $\frac{1}{2}\gamma (\Box\phi)^{2}$ to $(\ref{act1})$, equation (\ref{diffpot}) becomes \cite{cosmology}
\begin{equation}
\ddot{y}-\frac{3}{4}\frac{2c_{s}^{2}}{\gamma}V(t)y=0
\end{equation}
by using (\ref{sound}). We can define a new $\alpha$ in order to absorb this constant. Hence, let's define
\begin{equation}
\alpha^{\prime}=\frac{2c_{s}^{2}}{\gamma}\alpha
\label{newalpha}
\end{equation}
Therefore, $\alpha$  becomes $\alpha^{\prime}$ in potentials (\ref{exppot}) and (\ref{newpot}), and all the equations that are mentioned above that depend on $\alpha$. \\
Considering a plane wave perturbation $\propto e^{ikx}$, equation (\ref{0-i pert}) becomes,
\begin{equation}
\delta\ddot{\phi}_{k}+\frac{\dot{a}}{a}\delta\dot{\phi}_{k} +\Big(\frac{c^{2}_{s} k^{2}}{a^{2}}+\frac{\ddot{a}}{a}-\big(\frac{\dot{a}}{a}\big)^{2}\Big)\delta\phi_{k}=0
\end{equation}
By taking the limit of $t\longrightarrow \infty$ in (\ref{scalefactor}) is similar when taking the limit of the argument of Bessel's function to zero because of the decaying exponential function inside the argument of Bessel's functions. So for small $x$ , 
\begin{eqnarray}
 J_{0}(s)&\rightarrow& 1 \\
 Y_{0}(s)&\rightarrow& \frac{2}{\pi} \Big[\ln\big(\frac{s}{2}\big)+ 0.5772...\Big]
\end{eqnarray}
Hence, the scaling factor (\ref{scalefactor}) becomes as $t\longrightarrow \infty$,
\begin{equation}
a(t)=\Big[C_{1}+C_{2}\frac{2}{\pi}\Big(\ln\big(\frac{\sqrt{-3\alpha^{\prime}}}{2\kappa} e^{\frac{-\kappa t}{2}}\big)+0.5772 \Big)\Big]^{\frac{2}{3}}
\end{equation}
This equation can be expressed again as,
\begin{equation}
a(t)=[C_{1}+C_{2}\frac{2}{\pi}(\frac{-\kappa t}{2}+\beta)]^{\frac{2}{3}}=[C_{1}^{\prime}+C_{2}^{\prime}t]^{\frac{2}{3}}
\label{scalefactor_at_infinity}
\end{equation}
where $\beta= 0.5772+ \ln(\frac{\sqrt{-3\alpha^{\prime}}}{2\kappa})$ is just a constant, and $C_{1}^{\prime}=C_{1}+C_{2}\tfrac{2}{\pi}\beta$ and $C_{2}^{\prime}=-C_{2}\tfrac{\kappa}{\pi}$. By substituting (\ref{scalefactor_at_infinity}) in (\ref{0-i pert}), and solving the differential equation, we can get an idea about the evolution of $\delta\phi$ at a very large time-scale and for different wavelengths. For short wavelength perturbation $H$ and $\dot{H}$ are neglected because $\lambda_{ph}= \frac{a}{k}\ll c_{s} H^{-1}$,
\begin{equation}
\delta\phi\propto e^{\pm i c_{s}k t}
\end{equation} 
However, for long wavelength perturbation, the term $\frac{c^{2}_{s}}{a^{2}}\Delta\delta\phi$ is neglected because $\lambda_{ph}= \frac{a}{k}\gg c_{s} H^{-1}$; and hence, the solution to equation (\ref{0-i pert}) is
\begin{equation}
\delta\phi=D_{1}\pi+D_{2}\beta - D_{2}t\kappa
\label{perturbation} 
\end{equation}
Equation (\ref{perturbation}) can also be obtained by a second method; if we plug equation (\ref{scalefactor_at_infinity}) in (\ref{old pert1}), and choose $A \propto \kappa$ we would get equation (\ref{perturbation}) again. Note that the perturbation amplitude grows as a function of time only. \\
Action (\ref{act2}) to second order and integrating by parts yield
\begin{equation}
S=-\frac{1}{2}\int  d^{4}x\Big(\frac{\gamma}{c_{s}^{2}}\delta\phi^{\prime}\Delta\delta\phi^{\prime}+... \Big)
\end{equation}
 The canonically normalized quantum fluctuation variable \cite{Vikman} \cite{Mukhanov} is
\begin{equation}
v_{k} \sim \frac{\sqrt{\gamma}}{c_{s}} \: k \: \delta\phi_{k}
\end{equation}
with vacuum fluctuation
\begin{equation}
\delta v_{k} \sim \frac{1}{\sqrt{\omega_{k}}} \sim \frac{1}{\sqrt{c_{s}k}}
\end{equation}
and hence,
\begin{equation}
\delta\phi_{k} \sim \sqrt{\frac{c_{s}}{\gamma}} \: k^{-\frac{3}{2}}
\label{quantumpert}
\end{equation}
During inflation,
\begin{equation}
\frac{1}{a}\int a \: dt \simeq H^{-1}
\end{equation}
Matching long wavelength perturbations (\ref{old pert1}) with quantum perturbations (\ref{quantumpert})
\begin{equation}
A_{k} \sim \sqrt{\frac{c_{s}}{\gamma}} \: \frac{H_{c_{s}k\sim Ha}}{k^{3/2}}
\end{equation}
Hence, the gravitational potential in comoving scales $\lambda \sim 1/k$
\begin{equation}
\Phi_{\lambda} \sim A_{k}\; k^{3/2} \sim \sqrt{\frac{c_{s}}{\gamma}} \: H_{c_{s}k\sim Ha}
\label{gravitationalpot}
\end{equation}
In order to obtain the gravitational potential for comoving scales for potential (\ref{newpot}) from quantum perturbations, substitute (\ref{newscalefactor}), (\ref{newH}), and (\ref{alpha}) in (\ref{gravitationalpot}) with absolute value
\begin{equation}
\Phi_{\lambda} \sim \sqrt{\frac{c_{s}}{\gamma}} \; \times \sqrt{\frac{1}{3}}\;\frac{540(n+1)}{t_{i}^{n+1}} \; t^{n}|_{t:c_{s}k \sim Ha}
\end{equation}
Note that $\gamma$ is just a constant in the action (\ref{act2}). Hence, by choosing $n$, $\gamma$, and $t_{i}$ appropriately, one can fit the value of the gravitational potential to be equal to the measured value $\propto 10^{-5}$ in CMB experiments \cite{COBE}, \cite{WMAP}, and \cite{Planck 1}.
\section{Spectral Index Calculations}
In order to calculate the spectral index for potentials (\ref{exppot}) and (\ref{newpot}), we should calculate the slow-roll Hubble parameters,
\begin{eqnarray}
\varepsilon &\equiv& -\frac{\dot{H}}{H^{2}} \\
\eta &\equiv& \frac{\dot{\varepsilon}}{H\varepsilon} \\
n_{s}-1 &=& -2\varepsilon-\eta  \\
n_{t}&=&-2\varepsilon
\end{eqnarray}
For potential (\ref{exppot})
\begin{eqnarray}
\varepsilon=0 \\
\eta=0 \\
n_{s}=1 \\
n_{t}=0
\end{eqnarray}
It behaves like a cosmological constant and no gravitational waves. \\
For potential (\ref{newpot})
\begin{eqnarray}
\varepsilon &=& -\sqrt{\frac{3}{\alpha^{\prime}}}\frac{n}{t^{(n+1)}} \\
\eta&=&\sqrt{\frac{3}{\alpha^{\prime}}} \frac{(n+1)}{t^{(n+1)}}\\
n_{s}&=& 1+ \sqrt{\frac{3}{\alpha^{\prime}}}\frac{(n-1)}{t^{(n+1)}} \label{spectralindex}
\end{eqnarray}
Substituting (\ref{alpha}) and (\ref{newalpha}) in (\ref{spectralindex}) we obtain,
\begin{equation}
n_{s}\approx 1+\sqrt{\frac{3\gamma}{2c_{s}^{2}}} \frac{(n-1)}{540(n+1)}\Big(\frac{t_{i}}{t}\Big)^{(n+1)} \label{spectral2}
\end{equation}
where $t$ is evaluated at the horizon crossing $k=aH$. In order to make (\ref{spectral2}) less than one and match the data in CMB experiments \cite{COBE}, \cite{WMAP}, and \cite{Planck 1}, the second term must be negative. In potential (\ref{newpot}) inflation starts from $-\infty$; so $t_{i}$ and $t$ are negative. Hence, $n\in(-1,1)$ . 
The tensor spectral index
\begin{equation}
n_{t}\approx \sqrt{\frac{6\gamma}{c_{s}^{2}}} \frac{n}{540(n+1)}\Big(\frac{t_{i}}{t}\Big)^{(n+1)}
\end{equation}
If $n=0$, then there are no gravitational waves. 
\section{Conclusion}
In this paper, an exponential potential was substituted in the differential equation of MDM that relates any potential to any scaling factor in cosmology. At the limit of time goes to infinity, the density, pressure, and equation of state behave like dust in a matter-dominated universe, and in the limit of time goes to zero, a condition on the density can trigger inflation for some initial conditions satisfying the 60 e-folds condition. Another general potential is given that satisfies the 60 e-folds condition too. Furthermore, solutions to scalar perturbations are obtained for the general potential. This can be accomplished by taking the limit of $a(t)$ at infinity and substituting it in the $0-i$ perturbed Einstein's equation of a scalar field in the Newtonian gauge to get long wavelength perturbations. It is worth noting that after performing quantum perturbations, the obtained amplitude fluctuation from MDM can be tuned to be of the same order as the CMB. Finally, the spectral index for the mentioned potentials is calculated and the parameters were constrained. Hence, it was shown that mimetic inflation can have a red-tilt for the spectral index of adiabatic fluctuations. Therefore, MDM can have a model for dark matter, inflation with 60 e-folds at early times, and CMB's fluctuation.
\paragraph{\textbf{Acknowledgments}}
Many thanks to Prof. Ali Chamseddine for guidance and  Dr. Alexander Vikman for the illuminating discussions.

\end{document}